\documentclass[
 reprint,
superscriptaddress,
rsi,
amsmath,amssymb,
aps,
]{revtex4-1}

\usepackage{float}
\usepackage[utf8]{inputenc}
\usepackage{mathtools}
\usepackage{verbatim}
\usepackage{caption}
\usepackage{subcaption}
\usepackage{hyperref}
\usepackage{physics}

\usepackage{color}

\begin{document}
\title{Limitations in design and applications of ultra-small mode volume photonic crystals }

\author{Rubaiya Emran}
\thanks{Authors contributed equally to this work.}
\affiliation{Harvard College, Harvard University}

\author{Michelle Chalupnik}
\thanks{Authors contributed equally to this work.}
\affiliation{Department of Physics, Harvard University}
\affiliation{Currently at Aliro Quantum, Brighton}

\author{Erik N. Knall}
\affiliation{John A. Paulson School of Engineering and Applied Sciences, Harvard University}
\author{Ralf Riedinger}
\affiliation{Institut f\"ur Quantenphysik und Zentrum f\"ur Optische Quantentechnologien, Universit\"at Hamburg} 
\affiliation{The Hamburg Centre for Ultrafast Imaging}
\author{Cleaven Chia}
\affiliation{John A. Paulson School of Engineering and Applied Sciences, Harvard University}
\affiliation{Currently at Institute of Materials Research and Engineering, Agency for Science, Technology and Research Research Entities, Singapore}
\author{Marko Lon\v{c}ar}
\affiliation{John A. Paulson School of Engineering and Applied Sciences, Harvard University}

\begin{abstract}
    Ultra-small mode volume nanophotonic crystal cavities have been proposed as powerful tools for increasing coupling rates in cavity quantum electrodynamics systems. However, their adoption in quantum information applications remains elusive. 
    In this work, we investigate possible reasons why, and analyze the impact of different low mode volume resonator design choices on their utility 
    in quantum optics experiments. 
    We analyze band structure features and loss rates of low mode volume bowtie cavities in diamond and demonstrate independent design control over cavity-emitter coupling strength and loss rates. 
    Further, using silicon vacancy centers in diamond as exemplary emitters, we investigate the influence of placement imprecision.
    We find that the benefit on photon collection efficiency and indistinguishability is limited, while the fabrication complexity of ultra-small cavity designs increases substantially compared to conventional photonic crystals. 
    We conclude that ultra-small mode volume designs are primarily of interest for dispersive spin-photon interactions, 
    which are of great interest for future quantum networks. 
\end{abstract}

\maketitle

\section{Introduction}
The mode volume $V$ is an important figure of merit for an optical resonator as it indicates how tightly the resonator can confine the photon field. Since the emitter-cavity coupling $g$ is proportional to $1/\sqrt{V}$, small mode volumes can increase the cooperativity of nanophotonic cavity quantum electrodynamics (QED) systems, and even push them into the strong coupling regime. Large $g$ cavities are expected to be especially valuable for platforms aimed at indistinguishable single photon emission or robust, high-fidelity spin-photon gates, both of which are fundamental building blocks for quantum networks and other quantum information processing (QIP) systems \cite{duan2004, kimble2008}. To meet this need, ultra-small mode volume nanophotonic cavities (USMVC), with $V\ll \lambda^3/n^3$ where $\lambda$ is the wavelength of light, have been recently proposed \cite{Qimin_2009, Englund_LMV, chuck2019, yan2019, Kountouris:22, jingtong2024}. Enhancement of quantum emitters ranging from solid state emitters such as silicon-vacancy centers \cite{wein2018} to quantum dots \cite{Zhou:24} is possible via situation or implantation of the emitter within an USMVC. Moving beyond simulation, USMVCs have been experimentally realized in materials including silicon-on-insulator \cite{Weiss_SciAdv, majumdar}, indium phosphide \cite{Xiong2024}, and others.

One cavity QED platform that stands to benefit from USMVC designs is based on silicon-vacancy centers (SiVs) in diamond nanocavities. This well-established system allowed for demonstrations of various quantum photonics applications, including quantum network nodes \cite{Bhaskar_Nature, Nguyen_PRB, Nguyen_PRL, can_2023, bersin_telecom2024}, and indistinguishable single photon sources \cite{Knall2022}. 
Nominally, fidelities in all of these applications would substantially benefit from improved emitter-photon coupling.

In practical implementations, 
however, there are trade-offs that arise from stronger mode confinement.
For example, strong longitudinal confinement in a photonic crystal can impose non-adiabatic constraints on the mode, which can negatively impact its quality factor. Hence, an actual improvement in the emitter-cavity cooperativity, which scales as $Q/V$, is not immediately obvious. 
Therefore, design of USMVCs requires careful consideration of both transverse mode confinement and longitudinal mode confinement. We find that it is possible for strong transverse confinement to reduce mode volume without significantly lowering the photonic crystal quality factor. However, the dramatic enhancement promised by USMVCs is possible only if the quantum emitter is placed at the mode maximum. The extreme localization of the mode results in a rapid fall-off of the enhancement away from the mode maximum, placing
stringent requirements on the emitter-placement accuracy.  
For realistic estimations of the performance of USMVC designs, all of these trade-offs need to be accounted for.

While these considerations are generically applicable to all emitter-cavity platforms, here we focus on silicon-vacancy centers in diamond. 
In section \ref{sec:design}, we introduce a low mode volume photonic crystal design and demonstrate that strong transverse confinement does not substantially affect quality factor, allowing significant potential gains in cooperativity. Following this, in section \ref{sec:applications}, we evaluate the possible application both as a single photon source and as a robust high-fidelity spin photon interface, and discuss limitations on the maximum feasible enhancements from ultra-small mode volume cavities.

\section{Design of Low Mode Volume Cavities}
\label{sec:design}
In this section, we examine the capabilities and design limitations of USMV photonic crystal nanobeam cavities. We discuss the 
surprising robustness of USMVC quality factor to the tight confinement of the mode, as well as challenges associated with the practical implementation of USMVCs. Although here we present simulations for diamond photonic crystals with triangular cross-sections and with resonant wavelengths close to 737~nm, the design principles are general and can be used for photonic crystals constructed from other materials and resonance wavelengths.

\begin{figure}[ht]
\begin{subfigure}{\linewidth}
 \includegraphics[width=1\linewidth]{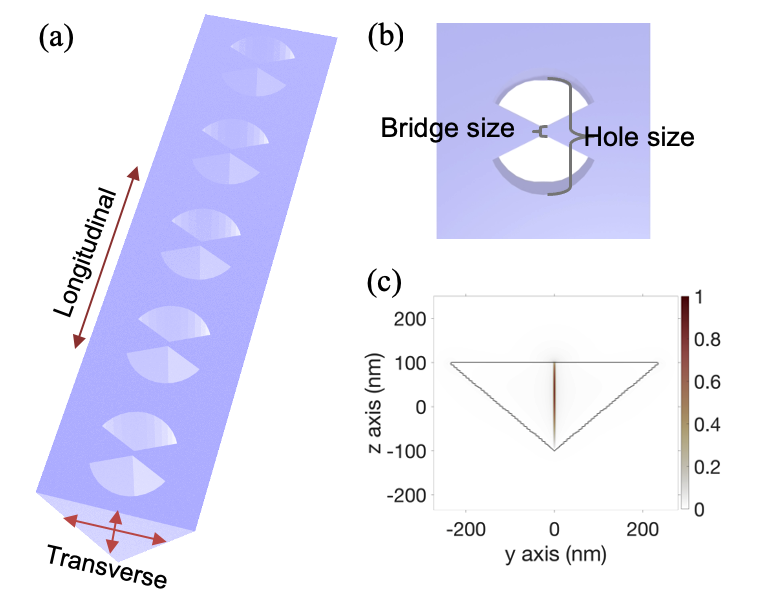}  
\end{subfigure}
\caption{(a) A rendering of the bowtie photonic crystal USMV cavity. (b) A unit cell of a bowtie photonic crystal cavity, with relevant geometric parameters labeled. (c) The simulated electric field intensity ($|E|^2$) of the triangular cross section at the center of a USMV photonic crystal cavity with a point-like bridge size of 0 nm.  
}
\label{fig:photonic-crystal-geometry}
\end{figure}

\begin{figure}[ht]
\centering
\includegraphics[width=8.5cm]{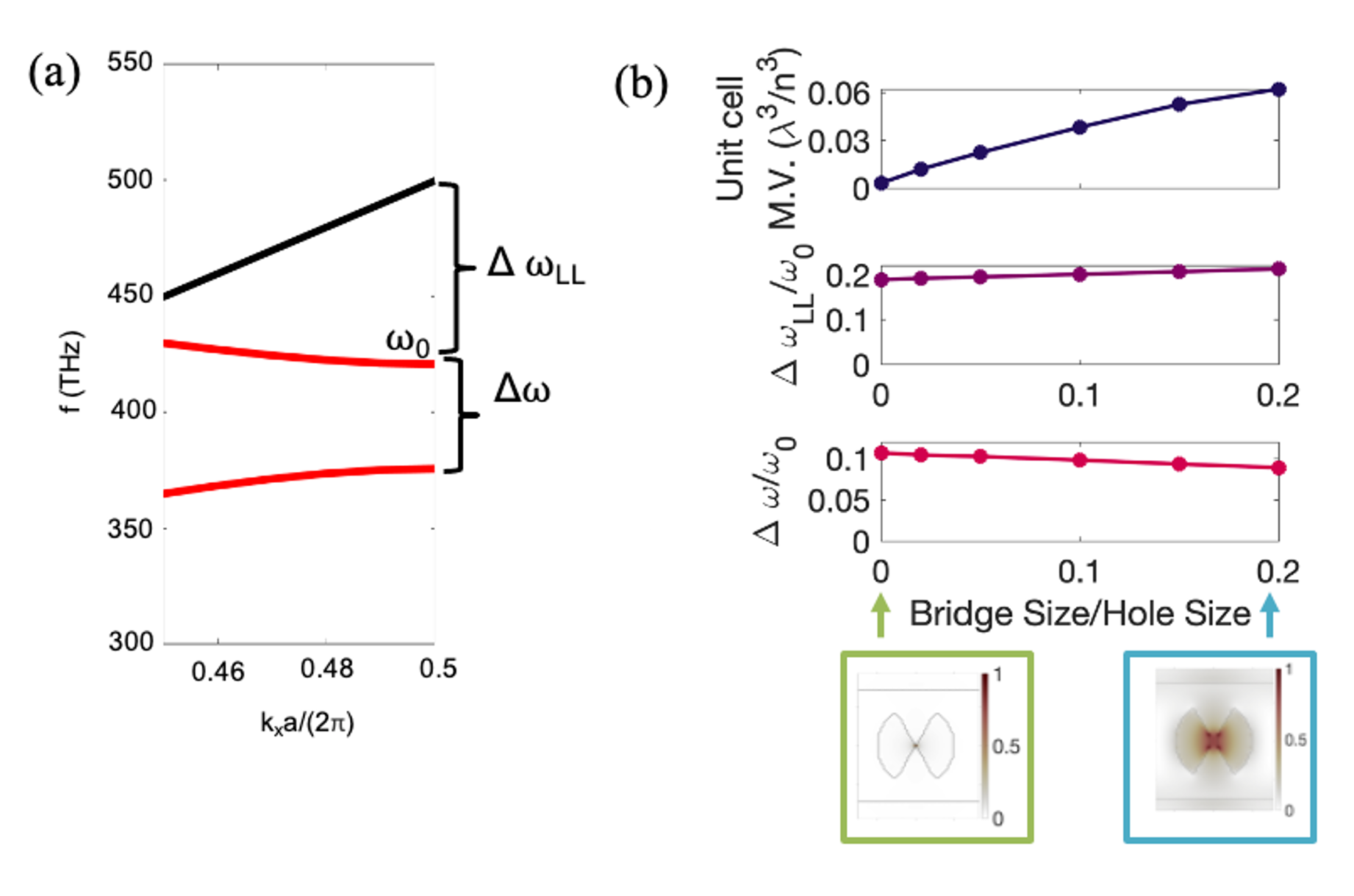}
\caption{ (a) The band structure for a bowtie unit cell photonic crystal, with resonant frequency of low mode volume mode labeled $\omega_0$, band gap labeled $\Delta \omega$, and mode distance to light line labeled $\Delta \omega_{LL}$. (b) The unit cell mode volume, distance from band to light line normalized by band frequency, and band gap normalized by band frequency for the low mode volume mode, plotted against bridge size to hole size ratio. Resonant frequencies are near $\lambda=737$~nm.}
\label{fig:unitcellMVs}
\end{figure}

As depicted in figure ~\ref{fig:photonic-crystal-geometry}, a photonic crystal nanobeam cavity is a quasi-periodic structure in which a one-dimensional (1D) chirped lattice of holes couples forward and backward propagating waves resulting in the formation of photonic band gaps. For frequencies in the band gap, the photonic crystals act as mirrors. A cavity is formed by the central chirped defect which shifts discrete modes from the band edge into the band gap of the mirror section. The mode volume of the photonic crystal, $V$, is given by total electric field energy divided by the maximum energy. 

\begin{align} 
     V = \frac{\int \epsilon(\vec{r}) \abs{E(\vec{r})}^2 d\vec{r}^3}{\max(\epsilon(\vec{r}) \abs{E(\vec{r})}^2)}
\end{align}

Photonic crystals with very low transverse mode volumes concentrated in the dielectric can be designed through consideration of electric field boundary conditions \cite{Weiss_SciAdv, Englund_LMV}. Electromagnetic boundary conditions require the normal component of the electric displacement field and the tangential component of the electric field to remain continuous between interfaces of varying permittivity. Through thoughtful design of unit cell cutout shapes, the ratio of the permittivity of air and the permittivity of the dielectric can be used to enhance the electric displacement field or the electric field within the dielectric. In one design method, dielectric and air slots are alternated across a central air hole to force the second order (air-like) mode into a small dielectric bridge.  As the number of slots and anti-slots in a single unit cell approaches infinity, the distribution of electric field becomes singular \cite{Weiss_design}. In this limit, the unit cell design resembles a bowtie, and mode volume is at a minimum. 

The low mode volume photonic crystals considered here are composed of prototypical unit cells with bowtie shaped holes (see figure \ref{fig:photonic-crystal-geometry}a-b) on a suspended diamond nanobeam with a triangular cross-section from angled etching \cite{burek2014}. The design concentrates the electric field intensity inside the center of the unit cell bridge for the relevant cavity mode (see figure \ref{fig:photonic-crystal-geometry}c). Our design consists of smaller bowties in the cavity defect with their diameter tapered down from the larger bowties of the outer mirror unit cells according to a third-degree polynomial. A well-established requirement for keeping photonic crystal intrinsic losses low is for the change in unit cell features between the mirror and cavity unit cells to be adiabatic \cite{quan2011}. We parameterize this adiabaticity by the number of unit cells in the tapered section between the inner cavity defect cell and outer mirror section. We chose to parameterize adiabaticity in this way following previous literature on elliptical hole-based photonic crystal nanobeam cavities [3]. In elliptical hole-based photonic crystal cavities which are scattering Q limited, mode volume increases linearly with the addition of extra unit cells in the section between the outer mirror cells and the inner cavity defect cell, given that outer mirror unit cell parameters and inner cavity defect unit cell parameters are tapered following a quadratic profile.

When considering contributions to mode volume in a one-dimensional photonic crystal, there are two directions along which the mode is confined. The longitudinal direction confines the mode parallel to the axis of the nanobeam, exploiting the band gap of photonic crystal structure. In the transverse direction, total internal reflection of the waveguide confines the mode along the two axes perpendicular to the nanobeam, with the confinement enhanced in-plane by the bowtie design. The transverse and longitudinal directions are indicated diagramatically over a bowtie photonic crystal cavity in figure \ref{fig:photonic-crystal-geometry}a. 

We first consider the effects of transverse mode confinement on band structure and loss rates. We simulate modes and photonic band structures (figure \ref{fig:unitcellMVs}) for a bowtie unit cell such that the second order TE-like mode is concentrated in the center of a dielectric bridge. The unit cells consist of a 1D diamond nanobeam with a triangular cross-section \cite{burek2012}. We use FDTD and MODE solvers (Lumerical) for our simulations. To create a metric for transverse confinement in unit cells, we introduce the concept of unit cell mode volume, which is defined identically to cavity mode volume except across a single unit cell with periodic boundary conditions, with the integration volume limited to this single unit cell.

Figure \ref{fig:unitcellMVs}a-b defines and plots key figures of merit of the unit cell performance for varying bridge-to-hole size ratios, indicating coupling strength (unit cell mode volume), transverse confinement (distance from light line), and longitudinal confinement (normalized band gap). The mode profiles for the two modes are also shown in figure \ref{fig:unitcellMVs}b. Unit cell mode volume increases as dielectric bridge size increases, as expected as the region concentrating light becomes larger. Simultaneously, the distance from light line for the low mode volume mode increases slightly while the band gap decreases slightly as the bridge size increases. At the X point in the band structure diagram, the distance of the low mode volume mode from the light line can be a measure of the quality factor: higher distance minimizes the mode overlap with the light cone (vacuum-like modes) and results in a higher intrinsic quality factor.
In principle, a tighter confinement requires a reduced bridge size and thus overlap of the optical mode with material. This brings the mode closer to the light line, which results in a lower intrinsic quality factor.
However, this is countered by a simultaneous increase of the bandgap with smaller bridge size (Fig. \ref{fig:unitcellMVs}b). Hence, the net effect is that a smaller bridge size creates greater transverse confinement (smaller mode volume) but has minimal influence
on scattering for a given number of mirror unit cells. 
The decrease in quality factor due to the smaller distance from the light line is small compared to other limits, such as fabrication imprecision.

\begin{figure*}[ht]
\includegraphics[width=16cm]{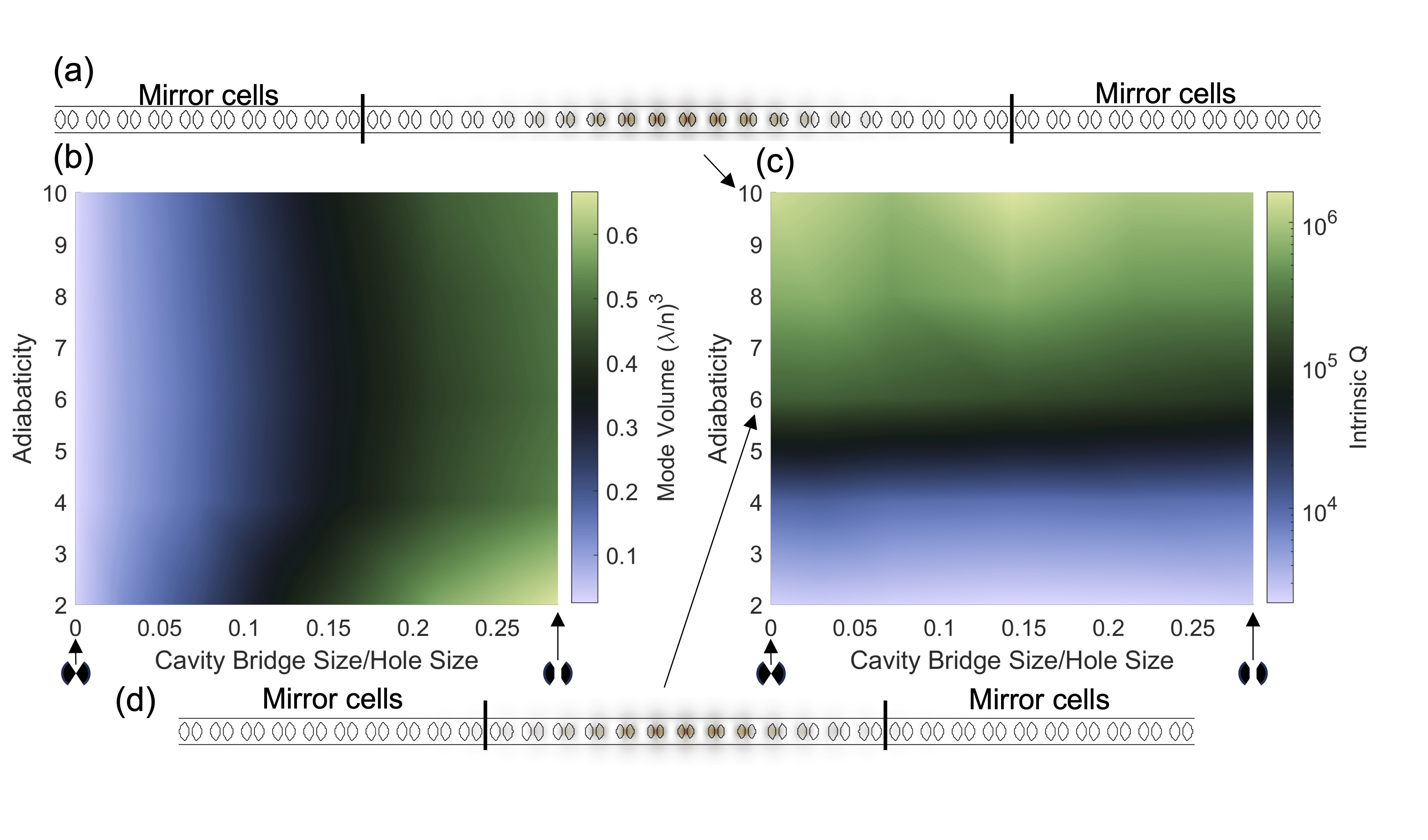}
\caption{ For full cavity simulations, effects of cavity bridge size and adiabaticity (number of taper cells) on intrinsic Q, and mode volume. (a) Longitudinal profile of electric field for adiabaticity of $10$, with bridge size of $20$ nm.
(b) Color plot of mode volume as a function of adiabaticity and bridge size. (c) Color plot of intrinsic Q as a function of adiabaticity and bridge size. (d) Longitudinal profile of electric field for adiabaticity of $6$, with bridge size of $20$ nm.
}
\label{fig:cavityMVs}
\end{figure*}

In contrast to transverse mode confinement, longitudinal mode confinement refers to mode confinement along the direction parallel to the waveguide. In theory, longitudinal mode confinement affects both cavity mode volume and intrinsic loss rates. We next consider these effects by simulating full cavities. Increasing the adiabaticity of the taper from mirror to cavity unit cell reduces intrinsic loss and increases quality factor. In this paper, we parameterize this adiabaticity $\alpha$ as the number of unit cells in the cubic taper between the innermost mirror cell and the center cavity unit cell (see figure  \ref{fig:cavityMVs}). 

Figure \ref{fig:cavityMVs}b shows the effect of adiabaticity and bridge size on cavity mode volume for ultra-low mode volume bowtie photonic crystals. Transverse confinement (parameterized here as cavity bridge size/hole size) has a larger effect on the total mode volume than the longitudinal confinement (parameterized here as adiabaticity). Decreasing bridge size improves transverse mode confinement, leading to the rapid decrease in mode volume shown in figure \ref{fig:cavityMVs}b. 
As adiabaticity changes, mode volume experiences only slight changes. With increasing adiabaticity, mode volume decreases slightly before gradually increasing again. The initial small decrease in mode volume as adiabaticity increases can be attributed to poor confinement in a leaky cavity, while the gradual increase in mode volume at higher adiabaticities shows the effect of worsening longitudinal confinement. These changes in mode volume are small, however. Figures \ref{fig:cavityMVs}a and \ref{fig:cavityMVs}d show the electric field intensity for two photonic crystal cavities of different adiabaticities but the same bridge size. The longitudinal confinement of the mode in figure \ref{fig:cavityMVs}d is tighter than the cavity in figure \ref{fig:cavityMVs}a, but the mode volume is relatively unaffected.

In contrast to cavity mode volume, the intrinsic cavity quality factor is strongly affected by longitudinal confinement (adiabaticity), as shown in figure \ref{fig:cavityMVs}c. For fixed bridge sizes, and for photonic crystals with very low adiabaticities, i.e., a taper length of only two unit cells, the intrinsic quality factor is low and the mode volume is high. As adiabaticity increases, intrinsic quality factor increases. 
However, in figure \ref{fig:cavityMVs}c, cavity bridge size has no strong influence on 
intrinsic quality factor. The weak influence of bridge size on the band structure shown in figure \ref{fig:unitcellMVs}b also reflects this fact.

In bowtie photonic crystal designs, the low mode volume arises from confining the air mode into a small dielectric bridge. As the dielectric bridge size decreases, the mode volume decreases while the air-to-dielectric ratio for the photonic crystal unit cell increases slightly. This effect results in larger band gaps that prevent greater scattering even with smaller distance to the light line and larger spread in $k$-space from confining the mode.

Because mode volume (intrinsic quality factor) is only weakly affected by varying longitudinal (transverse) confinement and is strongly affected by varying transverse (longitudinal) confinement, independent control of both mode volume and quality factor is possible. Notably, the intrinsic quality factor of low mode volume cavities will not necessarily be degraded, as long as the confinement is primarily transverse, rather than longitudinal. In sum, adiabaticity and bridge size essentially provide full control of quality factor and mode volume, allowing the cooperativity to be increased freely by reducing mode volume. 

For the cavities considered in figure \ref{fig:cavityMVs}, the mirror hole radius is
$100$ nm, the cavity hole radius 
$70$ nm, the lattice constant 
$280$ nm for the mirrors and $273$ nm for the cavity, and the beam width  
$475$ nm. The cavities have
a triangular cross section with a cross sectional angle of $50^\circ$. The ``best" cavity (i.e., with highest $Q/V$) considered in figure \ref{fig:cavityMVs}, which had an adiabaticity of $10$ and bridge size of $0$ nm (point-like contact), had a mode volume of $0.03$ $(\lambda/n)^3$ and intrinsic Q of $1.4 \times 10^6$. However, it is not currently feasible to make such small feature sizes in diamond in practice.

\begin{figure}[ht]
\includegraphics[width=7.5cm]{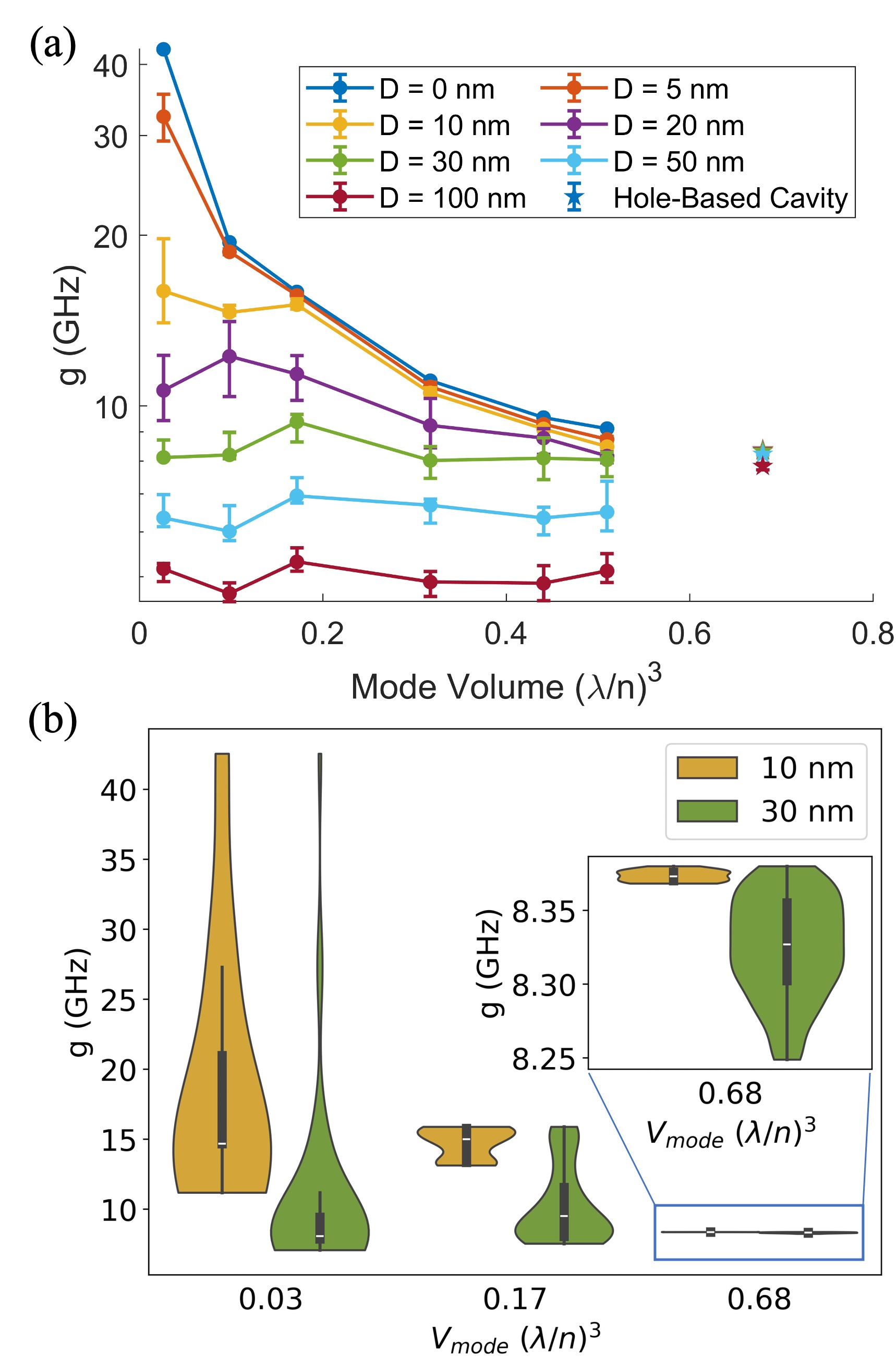}
\caption{ 
 (a) Relationship between ideal mode volume and average emitter-cavity coupling for different implantation accuracies, shown for six low mode volume cavities and one elliptical hole-based photonic cavity \cite{Nguyen_PRB}. The plotted values are the median value of $g$ over a region of diameter $D$ centered on the point within the dielectric at which $g$ is maximized.  Error bars show the range of 40th to 60th percentile in $g$. The value of the SiV dipole moment is calculated to be $\mu = 2.31$ Debye to be consistent with measured values of $g$ \cite{Nguyen_PRB}; alternative values for $\mu$ have been found in other work \cite{becker2017}. 
(b) Violin plot of
probability distribution of $g$ over implantation regions of diameter $10$ nm and $30$ nm for two low mode volume cavities ($V/(\lambda/n)^3=0.03, 0.17$) and one elliptical hole-based cavity ($V/(\lambda/n)^3=0.68)$.} 
\label{fig:effectiveMVs}
\end{figure}

For many applications, emitter-cavity coupling, or $g$, is a critical parameter, with a higher $g$ being desired (also see Appendix \ref{appendix:theory}). The cavity-emitter coupling $g$ is given by \cite{ReisererRempe}
\begin{align}
\label{eq:g}
   g =  \frac{| \vec{\mu}\cdot \vec{E}(\vec{r}) | }{ |\vec{E}_{\text{max}}|} \sqrt{\frac{ \omega}{2 \epsilon_0 \hbar V}}
\end{align}
where $\mu$ is the emitter dipole strength,  $\omega$ is the resonant frequency of the emitter, and $V$ is the mode volume of the cavity mode. In eq. \eqref{eq:g}, the electric field $E$ is evaluated at the location $\vec{r}$ where the emitter is located, which is not necessarily at the mode maximum $E_\text{max}$. An often neglected fact in the literature is that achieving the maximum calculated coupling strength $g$ for a given mode volume is dependent on the emitter being placed at the precise location of the electric field energy maximum. Imprecision and depth straggle of the ion implantation process used to embed SiVs can result in suboptimal coupling ($E(\vec{r}) \ll E_\text{max}$), which reduces the observed coupling. Other methods of vacancy-center placement including delta doping \cite{D1TC01538A}, in-situ annealing \cite{zuber2023}, manual placement of nanoparticles \cite{weinaliu2022}, and masked implantation \cite{Bhaskar_Nature} have similar limitations in precision. 
Figure \ref{fig:effectiveMVs}a plots the coupling strength $g$ against mode volume, given different implantation accuracies for an implantation region of diameter $D$ (excluding non-dielectric hole regions) in bowtie photonic crystal cavities with triangular cross-sections, and in an elliptical hole-based photonic crystal cavity \cite{burek2014} with a triangular cross-section. In figure \ref{fig:effectiveMVs}a, the coupling strengths are calculated as the median coupling strength over the implantation region. As variation in the vertical direction is typically negligible compared to variation in the lateral directions, the region considered is a circle centered on the point of maximum $g$ in the dielectric. Figure \ref{fig:effectiveMVs}b further illustrates the probability distribution of $g$ for the hole-based cavity and two selected bowtie cavities over implantation regions with diameters of $10$ nm and $30$ nm. The width of the shaded regions at each value of $g$ illustrate their relative weight in the probability distribution. In the center of each region is a box-and-whisker plot, with the box extending from the 25th to 75th percentile and the whiskers extending across 1.5 times the interquartile range. 

Figure \ref{fig:effectiveMVs}a shows $g$ quickly increases by an order of magnitude as implantation accuracy improves from an implantation uncertainty of diameter 100~nm to an implantation uncertainty of diameter 0~nm at a mode volume of 0.026 $\lambda^3/n^3$. For photonic crystal cavities with larger mode volumes, the drop in coupling occurs at larger implantation uncertainties. 
The couplings $g$ for a conventional photonic crystal cavity \cite{Nguyen_PRL} are shown as well, demonstrating that cavities based on elliptical holes are much less sensitive to implantation inaccuracy.
Figure \ref{fig:effectiveMVs}b shows that the tail of the distribution containing the maximal $g$ values becomes narrower for larger implantation uncertainties and lower mode volumes.
The squared edges in those distributions reflect the sharp cut-off at the edge of the circular region being considered.
For the lowest mode volume of $0.03$ $(\lambda/n)^3$ at an implantation uncertainty of $30$ nm, the probability of attaining the peak $g$ of $40$ GHz is extremely small, with the majority of the distribution weighted below $10$ GHz. This narrowing in distribution of maximal $g$ with increasing uncertainty corresponds to the rapid drop-off in median $g$ plotted in figure \ref{fig:effectiveMVs}a. By contrast, the probability distribution has negligible spread for the elliptical hole-based cavity, consistent with the low sensitivity to implantation inaccuracy shown in figure \ref{fig:effectiveMVs}a.
As state-of-the-art emitter placement is $D\geq30$~nm for SiVs, we find that currently conventional photonic crystals can on average outperform USMVCs. A bowtie design could probabilistically achieve substantially higher coupling, however, at lower device yield. Improved techniques for emitter implantation and precise fabrication of small features are needed to deterministically realize the simultaneous improvement in coupling and cooperativity from these designs.

\section{Applications}
\label{sec:applications}
In this section, we consider potential applications for USMVC in both the Purcell regime as well as the strong coupling regime, and we examine the extent of improvement USMVCs can provide in these applications. Appendix \ref{appendix:theory} gives further background on the applications described.

In the weak-coupling or Purcell regime, $\kappa \gg g \gg (\gamma, \gamma^*)$, and $C \gg 1$, where $\kappa$ is the cavity loss rate, $\gamma^*$ is the optical dephasing rate of the quantum emitter, $\gamma$ is the quantum emitter optical lifetime, and $C$ is the system cooperativity. In this regime, low mode volume cavities in diamond have applications as sources of indistinguishable single photons \cite{Knall2022}. The photon must escape the cavity at a faster rate than the rate at which it dephases or is scattered
by the emitter. We consider the case where the phonon sideband is spectrally well-separated from the zero-phonon-line and analyze the benefit of minimizing mode volume in this application. 

Here, we focus on two metrics: cavity efficiency and indistinguishability of photons. Cavity efficiency is defined as $\beta = \kappa \int^\infty_0 \langle a^\dag (t) a(t) \rangle dt$, with the creation and annihilation operators of the cavity field $a^\dagger, a$, and describes the efficiency by which a photon in the cavity will excite the emitter.
The indistinguishability \cite{wein2018}
\begin{align*}
     I = \frac{\int^\infty_0 \int^\infty_0 \abs{\langle a^\dag (t + \tau) a(t) \rangle}^2 dt d\tau}{\int^\infty_0 \int^\infty_0 \langle a^\dag (t + \tau) a(t + \tau) \rangle \langle a^\dag (t) a(t) \rangle dt d\tau}.
\end{align*}
describes the capability of two emitted photons to interfere.

\begin{figure}[ht]
\includegraphics[width=8cm]{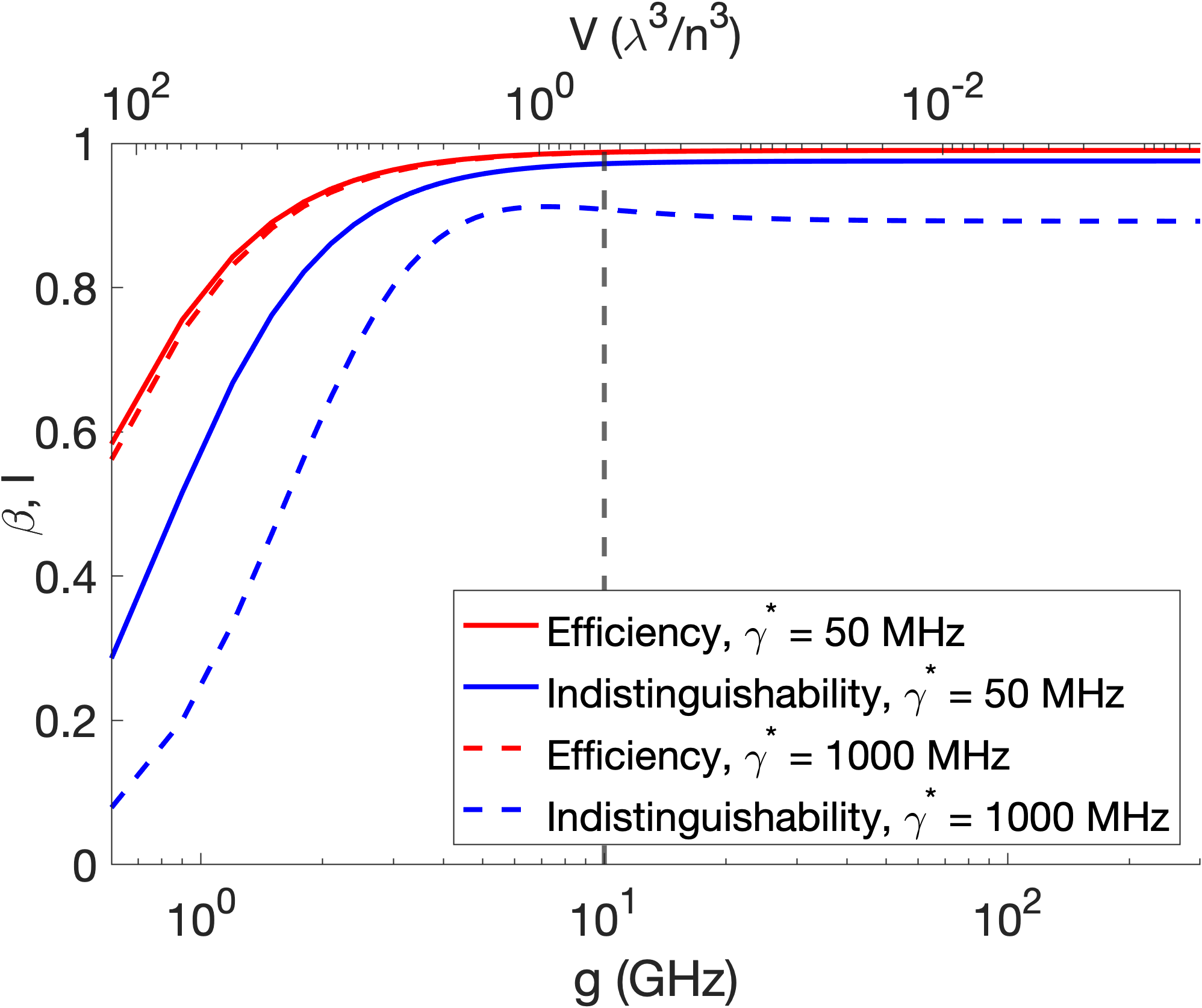}
\caption{Predicted indistinguishability $I$ (blue) and cavity efficiency (red) $\beta$ versus single photon-emitter coupling $g$ and respectively the 
corresponding mode volume for
a silicon-vacancy center with typical emitter-mode overlap in a diamond nanobeam photonic crystal \cite{Nguyen_PRB}.
For this system, $\kappa = \kappa_\textrm{wg} = 10$~GHz, $\gamma = $~100~MHz, $\gamma^* = $~50~MHz (solid line), resp. $\gamma^* = $~1~GHz. The vertical dashed 
line marks the parameters for conventional cavity \cite{Nguyen_PRB}.}
\label{fig:indist}
\end{figure}

Assuming a Markovian linewidth broadening process, in figure  \ref{fig:indist} we plot the indistinguishability and efficiency as a function of  $g$ (as well as  corresponding  mode volume for state of the art photonic crystal cavities) using the best reported $\kappa$ for a diamond nanobeam photonic crystal and using an experimentally typical emitter-mode overlap \cite{Nguyen_PRB}. The ultra-small features inherent in USMVC designs could decrease the scattering quality factor of fabricated structures. Our intent in simulating indistinguishability and cavity efficiency is to demonstrate any improvements USMVC can provide under realistic but optimal conditions. We therefore use the best reported $\kappa$ for elliptical hole-based diamond nanobeam photonic crystals as a ``best-case" loss rate. Silicon-vacancy center optical properties $\gamma$ and $\gamma^*$ at dilution refrigerator temperatures ($\sim$100 mK) are considered.  In this work, we assume optical properties of silicon-vacancy centers matching those observed of silicon-vacancy centers implanted in hole-based diamond photonic crystal nanobeams. A detailed analysis of the effects of USMVC designs on the optical properties of implanted silicon-vacancy centers is unfortunately beyond the scope of this work. 

Unlike in waveguide-emitter systems \cite{smith2017}, and as seen in figure \ref{fig:indist}, there is no tradeoff between efficiency and indistinguishability in cavity-emitter systems, as both follow the same trend, increasing with $g$. Efficiency and indistinguishability both increase with increasing $g$ (decreasing mode volume) until  $g \approx 5 $ GHz. After this point, efficiency and indistinguishability increase slowly, before reaching a plateau. 
Hence, reducing the mode volume below 1 $(\lambda / n)^3$ only yields minor improvements in photon extraction efficiency and indistinguishability in this system.
It appears that for strongly broadened emitters ($\gamma^*\gg 2\gamma$) the indistinguishability remains limited by the ratio of dephasing $\gamma^*$ over cavity bandwidth $\kappa$. 

\begin{figure}[ht]
\centering
\includegraphics[width=9cm]{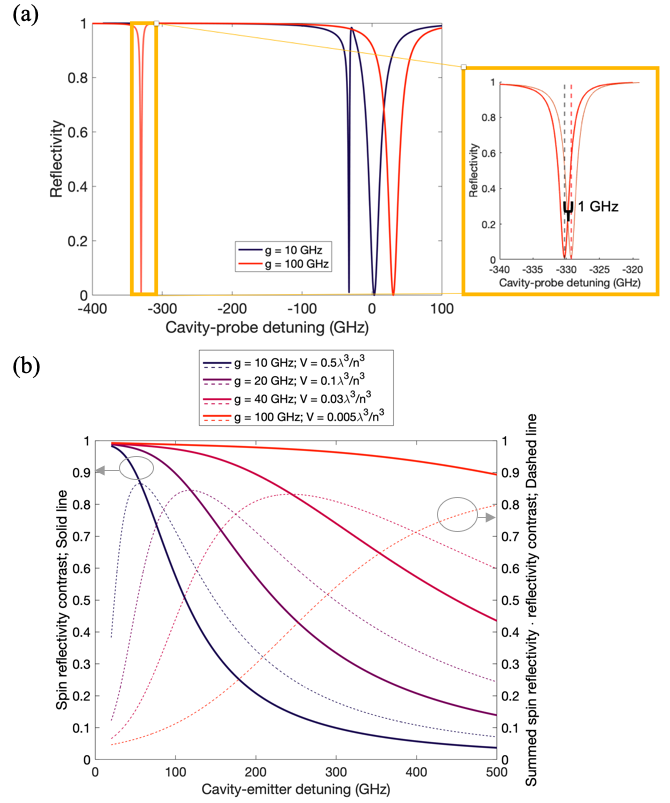}
\caption{(a) The simulated reflection spectra with $g = 10$ GHz corresponds to current achieved mode volumes (blue, $0.5$ $\lambda^3/n^3$), while the simulated reflection spectra with $g = 100$ GHz corresponds to a mode volume of $0.005$  $\lambda^3/n^3$ (red). 
For this system, $\kappa = 10$ GHz (quality factor $Q \approx 40,000$ ) and $\gamma$ = 100 MHz. The inset shows the differential optical Zeeman splitting (the difference between the optical transitions corresponding to the spin up and spin down states) of 1~GHz \cite{Bhaskar_Nature}. (b) The spin reflectivity contrast $|R_{\downarrow}-R_{\uparrow}|/(R_{\downarrow}+R_{\uparrow})$ (solid) and summed spin reflectivity multiplied by spin reflectivity contrast 
$ (R_{\downarrow}+ R_{\uparrow})|R_{\downarrow}-R_{\uparrow}|/(R_{\downarrow}+ R_{\uparrow})$ (dashed) with cavity-emitter detuning for different couplings and corresponding mode volumes, for a differential optical Zeeman splitting of 1 GHz. Simulations incorporate 50 MHz spectral drift for silicon-vacancy center ZPLs.
}
\label{fig:dispersive}
\end{figure}

In the strong coupling regime, where $g \gg \kappa, \gamma, \gamma^*$  ($C \gg 1$), USMVC could increase the fidelity of spin-photon gates. The electron spin of SiVs in photonic crystals have been used as memory qubits in quantum network nodes, which couple to photons via spin-dependent optical transitions. Of particular interest is spin-controlled switching of single photons \cite{Nguyen_PRL, Bhaskar_Nature, stas2022robust}, which requires high reflection amplitude contrast for the two spin states. This simultaneously allows for fast spin measurements, limiting the decoherence of coupled nuclear spin qubits \cite{stas2022robust}. 
Increased coupling $g$ allows for large spin-dependent contrast over a progressively larger cavity-emitter detuning range, reducing the system's sensitivity to the drifts and fluctuations. 
Furthermore, large detuning tolerance can allow for coarser tuning of cavities and an increased number of accessible emitters.

In figure \ref{fig:dispersive}a, we plot simulated cavity reflectivity for cavity with $g=10$~GHz or $100$~GHz, corresponding to state-of-the-art (blue) \cite{Bhaskar_Nature} and ultra-small mode volumes (red line), respectively. 
A spectral drift of 50 MHz is incorporated as a Gaussian convolution. The inset shows the shift of the reflection dip, equal to the differential optical Zeeman splitting, when the silicon-vacancy center changes spin state. In figure \ref{fig:dispersive}b, we plot the simulated contrast (and spin contrast multiplied by summed spin reflectivity, a metric which eliminates the normalization factor of summed reflectivity and may be a more useful metric in some experimental settings) for different couplings and corresponding mode volumes at different cavity-emitter detunings. The cavity detuning resulting in optimal contrast increases substantially for larger $g$, and simultaneously reduces sensitivity of the spin reflectivity contrast to the exact value of cavity-emitter detuning.
The spin-photon gate fidelities will be larger for the larger $g$ around a larger cavity-emitter detuning range, potentially adding tolerance for the inhomogeneous broadening of SiVs and for photonic crystal fabrication imperfections.

\section{Discussion}

In this paper, we investigated design criteria, sensitivity to fabrication imperfections, and possible applications of ultra-small mode volume photonic crystals in the context of cavity QED. Confining the mode along the transverse direction using a bowtie photonic crystal design can reduce mode volume without necessarily reducing quality factor.
While reducing the mode volume improves the cooperativity, the improvements to indistinguishability and collection efficiency of fluorescence photons from atom-like emitters are marginal
when compared to conventional photonic crystal resonators. For inhomogenously broadened emitters, indistinguishability cannot fully be restored in all cases. 
More sizable benefits of smaller mode volumes can be found in the dispersive regime, leading for example to a reduced sensitivity of spin-photon gates to emitter-cavity detuning. 

Investigating design criteria, we find that the mode volume of bowtie structures is dominated by the smallest possible feature size. 
The mode volume is minimized for a resonator that is just long enough to form a well-confined mode, here a defect of about 10 unit cells. 
A further increase in resonator-length has little influence on the mode volume, but can further increase the quality factor up to a nanofabrication-based limit, or until a target bandwidth is reached. 
Thus, intrinsic quality factor does not necessarily suffer in ultra-small mode volume photonic crystal designs. Intrinsic quality factor diminishes for ultra-small mode volume photonic crystals with strong longitudinal confinement, but it is virtually unaffected by changes in transverse confinement for bowtie photonic crystals. 

The promise of a photonic crystal with high quality factor and ultra-small mode volume, however, is tempered by fabrication and implantation precision constraints. Achieving ultra-small mode volumes in diamond photonic crystals with resonant wavelengths at silicon-vacancy center optical transitions requires fabrication of critical features at ultra-small sizes, as well as precise implantation to align emitters to the optical mode maximum. The requisite tolerance for emitter implantation error lies beyond current capabilites of scalable, ion-implantation-based methods. 
Hence, novel emitter placement and fabrication capabilities are required to unlock the potential of ultra-small mode volume cavities.

In conclusion, we find that the benefits of ultra-small mode volume photonic crystals to near-term experiments with narrow, atom-like solid state emitters are limited by current state of the art nanofabrication capabilities.
However, with fabrication advances, ultra-small mode volume cavities could be beneficial 
for dispersive spin-photon gates, as well as for broadband indistinguishable photon sources. These devices could contribute to scalable cavity QED platforms, especially through their reduced sensitivity to emitter-cavity detuning.

\section*{Methods}
The photonic crystal band structure and mode simulations were performed using Lumerical FDTD and MODE. The index of refraction of diamond used in our FDTD and MODE simulations was 2.40, which is equal to the index of refraction of diamond between the frequencies of 300-500 THz within  0.4 \%  \cite{permittivity}. This will introduce frequency uncertainties in our photonic band structures of around 0.4 \%. In FDTD simulations, we used finer spatial meshing of 2 nm on fine features of the bowtie photonic crystals, and coarser mesh on the features with larger critical dimensions. We also ran convergence tests with different sized spatial meshing to confirm that our mesh size was small enough to produce consistent results. In practice, imperfections during the fabrication process such as surface roughness or asymmetry can lower experimentally observed quality factors of fabricated photonic crystals from quality factors simulated with FDTD.
\\

\section*{Data availability statement}
The data that support the findings in this study are available from corresponding authors upon reasonable request. 

\section*{Acknowledgements}
The authors acknowledge helpful discussions with Bart Machielse, Kazuhiro Kuruma, Mikhail Lukin and Hyeongrak Choi. MC acknowledges support from the Department of Defense (DoD) through the National Defense Science and Engineering Graduate (NDSEG) Fellowship Program. ENK acknowledges that this material is based upon work supported by the National Science Foundation Graduate Research Fellowship under Grant No. DGE1745303. R. R. acknowledges support from the Cluster of Excellence `Advanced Imaging of Matter’ of the Deutsche Forschungsgemeinschaft (DFG) - EXC 2056 - project ID 390715994. CC acknowledges support from the Agency for Science, Technology and Research, Singapore. This work was supported in part by Army Research Office MURI (W911NF1810432), Office of Naval Research (N00014-20-1-2425), Airforce Office of Scientific Research (FA9550-20-1-0105), and AWS Center for Quantum Networking’s research alliance with the Harvard Quantum Initiative.

\appendix
\section{Theory of Single Photon-Emitter Coupling}
\label{appendix:theory}
Here, we review the theory behind the benefits of ultra-low mode volume photonic crystals for creating high cooperativity cavity QED systems. Mode volume and photonic crystal loss rates are included explicitly in the equations that govern cavity-QED systems. The dynamics of a silicon-vacancy center coupled to a single mode of a photonic crystal follow the master equation, 
\begin{align*} 
     \dot{\rho} &= - \frac{i}{\hbar} [H, \rho] + \kappa \left( a \rho a^\dag - \{a^\dag a, \rho\}/2 \right) \\
     & + \gamma^* \left( \sigma_+ \sigma_- \rho \sigma_+ \sigma_- - \{\sigma_+ \sigma_-, \rho \}/2 \right) \\
     & + \gamma \left( \sigma_- \rho  \sigma_+ - \{\sigma_+\sigma_-, \rho \}/2 \right),
\end{align*}
where $\rho$ is the density matrix, $\sigma_+$ and $\sigma_-$ are the two-level system lowering and raising operators, $a$ and $a^\dag$ are cavity mode ladder operators, $\kappa$ is the photonic crystal loss rate, $\gamma^*$ is the optical dephasing rate of the defect, and $\gamma$ is the silicon-vacancy center optical lifetime. (For the silicon-vacancy center optical transition over short times, $\gamma \approx 2 \gamma^*$). The interaction Hamiltonian after the rotating wave approximation is equal to 
\begin{align*} 
H = \hbar g (\sigma_- a^\dag + \sigma_+ a). 
\end{align*}

To obtain a useful figure of merit for a cavity QED system, the photon-emitter coupling can be compared to decoherence and loss rates, yielding the cooperativity $C = 4 g^2/( \kappa \gamma  )$. 
For many quantum applications, the strong cooperativity regime, with $C\gg 1$ is desired, such that the coupling rate is larger than the loss rates. For fixed $\gamma$, the cooperativity can be increased by decreasing $\kappa$, or increasing coupling $g$. For a cavity coupled to a waveguide, there are two cavity loss components: waveguide-cavity coupling loss ($\kappa_{wg}$) and intrinsic cavity loss typically limited by scattering  of light ($\kappa_{sc}$).  The  total loss is then given as $\kappa = \kappa_{wg} + \kappa_{sc}$.

Some applications (e.g., single photon sources) 
require high photon collection efficiencies through the waveguide, so high cooperativity and waveguide coupling are required simultaneously. In these cases, cavity designs that minimize mode volume can increase cooperativity while ensuring  $\kappa_{wg}>>\kappa_{sc}$. 
For other applications, it is desirable to achieve large quality factors (and small $\kappa$) in tandem with achieving small mode volumes, in particular for applications that require very high cooperativities, such as reflection amplitude readout of spin qubits.

\medskip
\bibliographystyle{unsrt}
\bibliography{sources}

\end{document}